\documentclass{article}
\usepackage{spconf,amsmath,graphicx}
\usepackage{epstopdf,booktabs,multicol,multirow,enumitem,footnote,type1cm,amssymb,calc}

\title{Non-Autoregressive Transformer ASR with CTC-Enhanced Decoder Input\vspace{-5pt}}
\name{
    \begin{tabular}{c}
        Xingchen Song$^{1,\dagger}$\thanks{$\dagger$ Work performed during an internship at Tencent AI Lab.},~Zhiyong Wu$^{1,2,\ddagger}$\thanks{$\ddagger$ Corresponding author.},~Yiheng Huang$^3$,~Chao Weng$^3$,~Dan Su$^3$,~Helen Meng$^{1,2}$
    \end{tabular}
    \vspace{-5pt}
}
\address{
    $^1$Tsinghua Shenzhen International Graduate School, Tsinghua University, Shenzhen\\~~
    $^2$The Chinese University of Hong Kong, Hong Kong\\
	$^3$Tencent AI Lab, Shenzhen\\
	sxc19@mails.tsinghua.edu.cn$^{\dagger}$,~~zywu@se.cuhk.edu.hk$^{\ddagger}$
	\vspace{-10pt}
}
\begin{document}
%
\maketitle
\begin{abstract}
\vspace{-1pt}
\textit{Non-autoregressive}~(NAR) transformer models have achieved significantly inference speedup but at the cost of inferior accuracy compared to \textit{autoregressive}~(AR) models in automatic speech recognition~(ASR). Most of the NAR transformers take a fixed-length sequence filled with MASK tokens or a redundant sequence copied from encoder states as decoder input, 
they cannot provide efficient target-side information thus leading to accuracy degradation. To address this problem, we propose a CTC-enhanced NAR transformer, which generates target sequence by refining predictions of the CTC module.
Experimental results show that our method outperforms all previous NAR counterparts and achieves \textbf{50x} faster decoding speed than a strong AR baseline with only $0.0\sim0.3$ \textbf{absolute} CER degradation on Aishell-1 and Aishell-2 datasets. 
\end{abstract}
\begin{keywords}
\vspace{-1pt}
non-autoregressive, transformer, ctc
\end{keywords}
\vspace{-3pt}
\section{Introduction}
\label{sec:intro}
\vspace{-3pt}
Recently, the self-attention based encoder-decoder framework, called transformer~\cite{ic18/speechtransformer}, has achieved very promising results for automatic speech recognition~\cite{asru19/compare_rnn_transformer} when comparing to traditional hybrid models~\cite{asru11/kaldi} and CTC based models~\cite{icml06/ctc}.
However, such model suffers from a high latency during the inference process as it translates a source sequence in an \textit{autoregressive} manner, i.e.(see Fig.\ref{fig:model}(a)), it generates a target sentence character by character from left to right and the generation of $t$-th token $y_t$ depends on previously generated tokens $y_{1:t-1}$ and encoded audio representations $\mathcal{E}(x)$:
\begin{equation}
    \label{eq:ar}
    \vspace{-2pt}
    y_t = \mathcal{D}(y_{1:t-1}, \mathcal{E}(x))
    \vspace{-2pt}
\end{equation}
where $\mathcal{E}(\cdot)$ and $\mathcal{D}(\cdot)$ denote the encoder and decoder part of the model respectively, $x$ is the input audio features and $\mathcal{E}(x)$ is the output of encoder, i.e., a series of encoded hidden states at the top layer of the encoder.

To speed up the inference of speech recognition, \textit{non-autoregressive} transformers have been proposed~\cite{arxiv/laso, arxiv/spike, arxiv/maskctc, arxiv/maskpredict}, which generate all target tokens simultaneously~\cite{arxiv/laso, arxiv/spike} or iteratively~\cite{arxiv/maskctc, arxiv/maskpredict}. We notice that the encoder of AR transformers and that of NAR transformers are the same thus the differences lie in the decoder. More specifically, instead of using previously generated tokens as in AR decoders, NAR decoders take other global signals derived from either empirical statistics~\cite{arxiv/laso, arxiv/maskpredict} or encoded speech sequence~\cite{arxiv/spike}. 

For example, the NAR decoders in \textit{listen attentively and spell once} (LASO) \cite{arxiv/laso} and \textit{listen and fill in missing letter} (LFML)~\cite{arxiv/maskpredict} take fixed-length sequence filled with MASK tokens as input to predict target sequence:
\begin{equation}
    \label{eq:laso}
    \vspace{-1pt}
    y_{t} = \mathcal{D}(MASK_{1:L}, \mathcal{E}(x))
    \vspace{-1pt}
\end{equation}
where 
$L$ is a constant number that was predefined and used for all sentences. The settings of this predefined length is usually based on empirical analysis of durations of audio and the final output length is determined by the position where the first end-of-sequence~(EOS) token appears. To guarantee the performance of the model, the predefined length $L$ is often much longer than the actual length of the target sequence $T$. This results in extra calculation cost and affects the inference speed. 
In order to estimate the length of the target sequence accurately and accelerate the inference speed, ~\cite{arxiv/spike} proposed a \textit{spike-triggered non-autoregressive}~(ST-NAR) model, which introduces a CTC module~\cite{icml06/ctc} to predict the target length and takes a copy of encoder hidden states $\hat{\mathcal{E}}(x)$ as the decoder input, and the copy process is guided by CTC spikes, which represent positions of the state that should be copied. After that all target tokens are simultaneously predicted:
\begin{equation}
    \label{eq:spike}
    \vspace{-1pt}
    y_{t} = \mathcal{D}(\hat{\mathcal{E}}(x), \mathcal{E}(x))
    \vspace{-1pt}
\end{equation}
where $\hat{\mathcal{E}}(x) = (\mathcal{E}(x)_{p_1}, \mathcal{E}(x)_{p_2}, ..., \mathcal{E}(x)_{p_{T'}})$ are the copied encoder states. $p_i$ is the position where CTC spike occurs and $T'$ is the number of CTC spikes, it can be seen as the target length predicted by CTC.

While ST-NAR models significantly reduce the inference latency, they suffer from large accuracy degradation compared with their AR counterparts. We notice that in AR models, the generation of $t$-th token $y_t$ is conditioned on previously generated tokens $y_{1:t-1}$, which provides target side context information. In contrast, ST-NAR~\cite{arxiv/spike} models generate tokens conditioned on the copied encoder states $\hat{\mathcal{E}}(x)$ guided by spikes, such information is redundant and indirect because the copied states are still in the domain of audio features.

Consequently, the decoder of ST-NAR model has to handle the recognition task conditioned on less and weaker information compared with its AR counterparts, thus leading to inferior accuracy. In this paper, we aim to enhance the decoder inputs of NAR models so as to reduce the difficulty of the task that the decoder needs to handle. Our basic idea is to directly feed greedy CTC predictions $\hat{y} = CTC(\mathcal{E}(x))$ to the decoder. We conjecture that the CTC outputs can provide more closed information to the target-side tokens $y$, thus reducing the gap between AR and NAR models. Experimental results on two corpora demonstrate that our model outperforms all compared NAR models and exceeds strong AR baseline by \textbf{50x} faster speed with comparable performance.
\vspace{-5pt}
\section{Methodology}
\label{sec:model}
\vspace{-2pt}
\subsection{Related Work}
\begin{figure}[htp]
    \centering
    \vspace{-5pt}
    \includegraphics[scale=0.4]{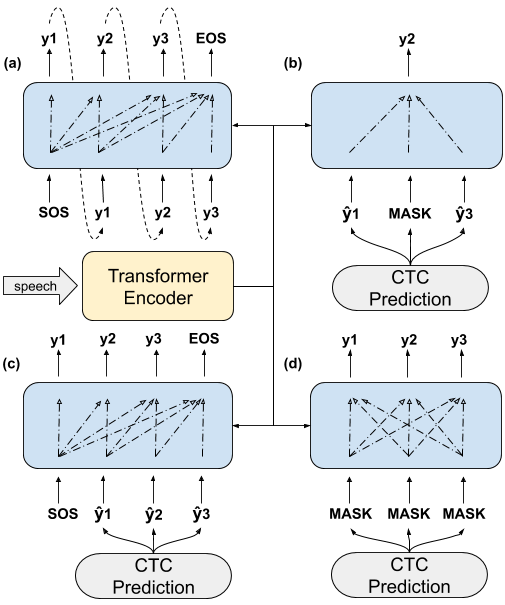}
    \vspace{-5pt}
    \caption{Comparison between AR and NAR transformer \textbf{during inference}. The blue boxes indicate Transformer Decoder and the dash-dotted lines inside them indicate the dependency between tokens. (a) AR Transformer. The dashed line limits the emission of $y_t$ by waiting for the end of generating $y_{t-1}$. (b) MP-CTC~\cite{arxiv/maskctc}. (c) proposed NAR Transformer. (d) upgraded version of LASO~\cite{arxiv/laso}.}
    \label{fig:model}
    \vspace{-5pt}
\end{figure}

Recently,~\cite{arxiv/maskctc} trains a Transformer encoder-decoder model with both Mask-Predict~\cite{emnlp19/mask-predict} and CTC objectives~(MP-CTC). As shown in Fig.\ref{fig:model}(b), during inference, the input of decoder is initialized with the greedy CTC prediction $\hat{y}$ and low-confidence tokens are masked based on the CTC probabilities. The masked low-confidence tokens $y_{l}$ are then predicted by the decoder conditioning on the high-confidence tokens $\hat{y}_{h}$:
\begin{equation}
    \label{eq:mp-ctc}
    y_{l} = \mathcal{D}(\hat{y}_{h}, \mathcal{E}(x))
\end{equation}

\vspace{-5pt}

While this approach use CTC predictions as decoder input as we do, the Mask-Predict~\cite{emnlp19/mask-predict} manner that was originally proposed for neural machine translation~(NMT) task may not be suitable for ASR as it suffers from two drawbacks:~1)~As pointed out in~\cite{arxiv/semi-mask-predict}, the observed tokens (high-confidence CTC predictions) are \textit{not} always correct and the masked tokens (low-confidence CTC predictions) are \textit{not} always wrong, this may further increase mis-recognitions when re-decoding low-confidence but correct tokens to wrong tokens or making predictions based on high-confidence but incorrect tokens.~2)~Previous works~\cite{acl20/study-of-nar, arxiv/hardenc_easydec} show that the decoder is more sensitive to target-side information than source-side information and ASR is more serious than NMT due to the different characteristics between the input acoustic signals and the output linguistic symbols~\cite{acl20/study-of-nar}. This suggests that given the full context of audio features $\mathcal{E}(x)$ and the full context of characters $\hat{y}_h$ as in~Eq.\ref{eq:mp-ctc}, the mask-based character prediction can be easily filled out only with the target context without using any speech input~(i.e., the network only learns the decoder). In this situation, the training-inference discrepancy, where the ground truth is used as decoder input during training and replaced by CTC prediction during inference, will largely affect the performance and this is verified by our study in Fig.\ref{fig:rubost}.
\vspace{-2pt}
\subsection{Proposed Method}
As shown in Fig~.\ref{fig:model}(c), we address the above problems by limiting the target context to the past thus forcing the model to pay more attention to source-side context $\mathcal{E}(x)$:
\begin{equation}
    \label{eq:cm}
    y_{t} = \mathcal{D}(\hat{y}_{1:t-1}, \mathcal{E}(x))
\end{equation}

It is interesting to note that Eq.\ref{eq:cm} is similar to its AR baseline described in Eq.\ref{eq:ar}. The difference is that we can get all target tokens $\hat{y}_{1:T'}$ before the decoder starts to translate while AR model cannot. This allows us to make the decoding process fully paralleled while maintaining the AR properties for each generated token and this is the reason that we can achieve \textbf{50x} speedup while capturing the performance of the AR counterpart.

Here, we propose two concrete methods to train transformer model to support the decoding type described in Eq.\ref{eq:cm}:
\begin{itemize}
  \item[1.] \textit{Teacher Forcing~(CM)}: we feed ground truth to the decoder during training and CTC prediction during inference. CM means we use Causal Mask (CM) as decoder attention mask to discard target-side future context to obey the left-to-right rule.
  \item[2.] \textit{CTC Sampling~(CM)}: during training, we perform on-the-fly CTC decoding to get predicted length $T'$ and adjust the decoder input according to it. Specifically, if $|T - T'| <= 2$, we use CTC prediction $\hat{y}$, otherwise we use ground truth as input. 
\end{itemize}

\subsection{Variations for Comparison}
\vspace{-5pt}
Beside the Causal Mask, we also replace it with Padding Mask (PM), which only masks the padded tokens in a minibatch and is also used in MP-CTC~\cite{arxiv/maskctc}, for \textit{CTC Sampling} to remove the context restrictions to validate our conjecture that the full context of targets may lead to performance degradation due to the sensitivity of decoder towards target-side information. This can be formalized as:
\begin{equation}
    \label{eq:pm}
    y_{t} = \mathcal{D}(\hat{y}_{1:T'}, \mathcal{E}(x))
\end{equation}

To make a thorough comparison, we also proposed an upgraded version of LASO~\cite{arxiv/laso}. As shown in Fig.\ref{fig:model}(d) and Eq.\ref{eq:bf-pm}, we replace the predefined length $L$ in Eq.\ref{eq:laso} with CTC predicted length $T'$:
\begin{equation}
    \label{eq:bf-pm}
    y_{t} = \mathcal{D}(MASK_{1:T'}, \mathcal{E}(x))
\end{equation}

During training, ground truth is used as decoder input and all target tokens are replaced with MASK. In this case, target-side information only contains the length information and the semantic context can be retrieved only from the source-side information $\mathcal{E}(x)$~\cite{arxiv/laso}. We name this strategy as \textit{Mask Forcing~(PM)}. Note that during inference, the length information will be provided by CTC.
\vspace{-5pt}

\section{Experiments}
\vspace{-5pt}
\label{sec:exp}
In this work, all experiments are conducted using ESPnet tranformer~\cite{is18/espnet}$\footnotemark[1]$ with CTC joint training~\cite{asru19/compare_rnn_transformer}. The model and experiment setups are the same as~\cite{asru19/compare_rnn_transformer} except that we use SpecAugment~\cite{is19/specmask} and set $total\_training\_epoch=200$ for all experiments. Due to space limitations, the details are omitted and interested readers can refer to ~\cite{asru19/compare_rnn_transformer}.
\vspace{-10pt}
\subsection{178-hours Corpus: AISHELL-1}
\begin{figure}[!htp]
    \centering
    \vspace{-5pt}
    \includegraphics[scale=0.3]{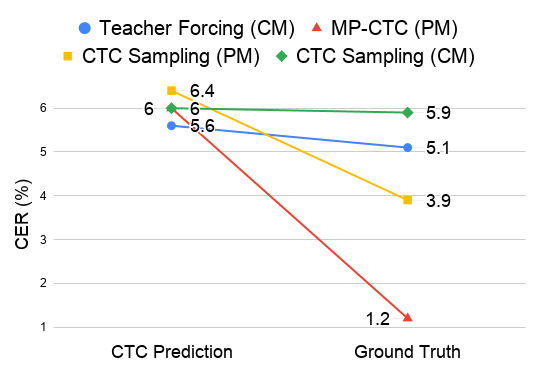}
    \vspace{-5pt}
    \caption{The CER drop when change the decoder input. Note that for MP-CTC~\cite{arxiv/maskctc}, the ground truth is obtained by replacing the non-masked tokens in CTC prediction with the corresponding correct tokens.}
    \vspace{-5pt}
    \label{fig:rubost}
\end{figure}

\vspace{-5pt}
\label{sec:aishell1}
We first conduct experiments and analyses on a public mandarin speech corpus AISHELL-1~\cite{cocosda17/aishell-1}. As shown in Table~\ref{tab:aishell-1}, our best strategy speeds up the AR baseline (beam=10) by \textbf{50x} times with only $0.0\sim0.2$ absolute CER degradation and achieves \textbf{5x} faster speed than the greedy result (beam=1) with even better CER. We find that both \textit{Teacher Forcing (CM)} and \textit{MP-CTC~(PM)} perform better then \textit{Mask Forcing~(PM)}, this is partly because the latter one cannot provide efficient target-side information.
\begin{table}[!htp]
    \centering
    \caption{The character error rate (CER) of the systems on AISHELL-1. Real-time factor (RTF) is computed as the ratio of the total inference time to the total duration of the dev set. Inference is done with $batch size=8$, on an NVIDIA Tesla P40 GPU.}
    \vspace{2pt}
    \scalebox{0.75}{
        \begin{tabular}{c|c|c|c|c}
            \toprule[1.5pt]
            Training Strategy & Decode Type  & Beam & Dev / Test & RTF \\
            \midrule
            \multicolumn{5}{c}{\textit{Autoregressive Transformer}} \\
            \midrule
            Baseline~\cite{asru19/compare_rnn_transformer} (with & Fig.\ref{fig:model}(a) \& Eq.\ref{eq:ar} & 1  & 5.6 / 6.6 & 0.0186\\
            CTC Joint Training) & & 10 &  \textbf{5.6} / \textbf{6.1} & 0.1703\\
            \midrule
            \multicolumn{5}{c}{\textit{Non-autoregressive Transformer}} \\
            \midrule
            CTC Sampling~(PM) 
                    & ~~~~-~-~~~~~ \& Eq.\ref{eq:pm} & - & 6.4 / 7.2 & 0.0037\\
            CTC Sampling~(CM) 
                    & Fig.\ref{fig:model}(c) \& Eq.\ref{eq:cm} & - & 6.0 / 6.8 & 0.0037\\
            Mask Forcing~(PM) 
                    & Fig.\ref{fig:model}(d) \& Eq.\ref{eq:bf-pm} & - & 6.0 / 6.8 & 0.0037\\
            MP-CTC~(PM)~\cite{arxiv/maskctc}~$\footnotemark[2]$
                    & Fig.\ref{fig:model}(b) \& Eq.\ref{eq:mp-ctc} & - & 6.0 / 6.6 & 0.0134\\
            Teacher Forcing~(CM) 
                    & Fig.\ref{fig:model}(c) \& Eq.\ref{eq:cm} & - & \textbf{5.6} / \textbf{6.3} & \textbf{0.0037}\\
            \bottomrule[1.5pt]
        \end{tabular}
    }
    \vspace{-10pt}
    \label{tab:aishell-1}
\end{table}
\footnotetext[1]{~~https://github.com/espnet/espnet} 
\footnotetext[2]{~We re-implement MP-CTC~\cite{arxiv/maskctc} using official code retrieved from https://github.com/espnet/espnet/pull/2070 and set the number of total decoding iterations to $5$ as suggested in \cite{arxiv/maskctc}.}
Besides, we also find that strategies with CM always outperform those with PM, i.e., \textit{Teacher Forcing~(CM)} and \textit{CTC Sampling~(CM)} work better than \textit{MP-CTC~(PM)} and \textit{CTC Sampling~(PM)}, respectively. This suggests that given the full context of speech features $\mathcal{E}(x)$, target-side context should be limited to avoid excessive reliance on it. Such reliance is caused by the aforementioned sensitivity of decoder and we validate it in Fig.\ref{fig:rubost}, where we change the decoder input during inference to see the performance gap between ideal and reality, i.e., feed ground truth or CTC prediction to the decoder respectively. By comparing \textit{CTC Sampling (CM)} and \textit{CTC Sampling (PM)} in Fig.\ref{fig:rubost}, we can clearly see that PM based decoder is more sensitive to the input. As for \textit{MP-CTC (PM)} and \textit{CTC Sampling (PM)}, although both of them use PM as attention mask, the performance of the former one is more affected by the input. This is because the decoder of \textit{MP-CTC (PM)} can only see ground truth during training thus suffers more from training-inference discrepancy. In contrast, \textit{Teacher Forcing~(CM)}, which limits the target-side context to obey the autoregressive manner, performs more stable and much closer to the upper bound of the trained model, indicating the effectiveness of our proposed method and we will use it for all our following experiments.

\begin{figure}[!htp]
    \centering
    \vspace{-5pt}
    \includegraphics[scale=0.27]{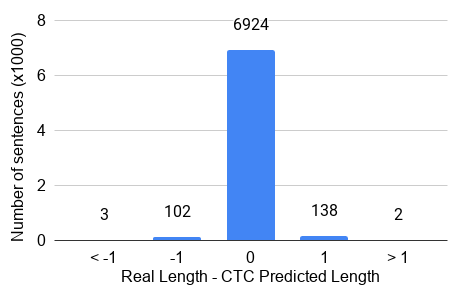}
    \vspace{-10pt}
    \caption{The analysis of the predicted length on test set. The histogram record the difference between the target length $T$ and the CTC predicted length $T'$. 
    It will not cause irreversible effects when the value is less then or equal to zero, as the decoder is still able to correct the CTC output by predicting EOS token at the end of sentence.
    }
    \vspace{-15pt}
    \label{fig:len}
\end{figure}

To study how CTC contributes to our NAR models, we analyze the predicted length by CTC. As shown in Fig.\ref{fig:len}, the accuracy of length prediction is higher than 96\%, which means that $T'$ is accurate enough to lower the computation cost when we upgrade Eq.\ref{eq:laso} to Eq.\ref{eq:bf-pm}

Besides, we also provide a comparison between CTC prediction and NAR Decoder prediction in Table~\ref{tab:ctcVSdecoder}. Due to the strong conditional independence assumption, CTC prediction has a poor performance. However, the decoder can still correct the input  based on both speech context and partial target context, indicating that the target-side information provided by CTC is useful and efficient for our NAR decoder.
\begin{table}[!htp]
    \centering
    \vspace{-10pt}
    \caption{CER comparison between CTC prediction and decoder prediction for proposed NAR transformer.}
    \vspace{2pt}
    \scalebox{0.7}{
        \begin{tabular}{c|c}
        \toprule[1.5pt]
        Result & Dev / Test \\
        \midrule
        CTC Prediction & 6.0 / 6.7\\
        NAR Decoder Prediction & \textbf{5.6} / \textbf{6.3}\\
        \bottomrule[1.5pt]
        \end{tabular}
    }
    \vspace{-5pt}
    \label{tab:ctcVSdecoder}
\end{table}


Lastly, we give a CER comparison with other approaches in Table~\ref{tab:aishell-1-compare}. To make a fair comparison, we also add Speed-Perturb~\cite{is15/speed_perturb} based augmentation to our training and approximate the number of parameters based on the description in the previous studies. It can be seen that our strong AR baseline reaches the state-of-the-art performance as it outperforms all previously published results and our proposed NAR model also beats all previous NAR counterparts. Considering the \textbf{50x} faster decoding speed, the gap between AR and NAR Transformer is tolerable.
\begin{table}[!htp]
    \centering
    \vspace{-10pt}
    \caption{CER comparison with Hybrid and End-to-End models on Aishell-1~\cite{cocosda17/aishell-1}. We use beam=10 for our AR model and keep it unchanged in our following experiments.
            $\blacktriangle$: SpecAugment~\cite{is19/specmask} is used.
            $\blacklozenge$: Speed-Perturb~\cite{is15/speed_perturb} is used.}
    \vspace{2pt}
    \scalebox{0.75}{
        \begin{tabular}{c|c|c|c}
            \toprule[1.5pt]
            Model  & Dev / Test & RTF & Params \\
            \midrule
            \multicolumn{4}{c}{\textit{End-to-End Autoregressive}} \\
            \midrule
            LAS~\cite{ic19/component_fusion} & -~~/ 8.7 & - & $\approx$ 156 M\\
            Transformer~\cite{arxiv/laso} & 6.1 / 6.6 & - & $\approx$ 58 M\\
            \midrule
            \multicolumn{4}{c}{\textit{End-to-End Non-autoregressive}} \\
            \midrule
            CTC~\cite{arxiv/spike}~$\blacktriangle$ & 7.8 / 8.7 & - & - \\
            ST-NAR~\cite{arxiv/spike}~$\blacktriangle$ & 6.9 / 7.7 & - & $\approx$ 31 M \\
            MP-CTC~\cite{arxiv/maskctc}~$\blacktriangle$~$\footnotemark[2]$ & 6.0 / 6.7 & 0.0134 & 29.7 M \\
            LASO(big)~\cite{arxiv/laso}~$\blacktriangle\blacklozenge$ & 5.8 / 6.4 & - & $\approx$ 105 M\\
            \midrule
            \multicolumn{4}{c}{\textit{Traditional Hybrid}} \\
            \midrule
            Kaldi/nnet3~\cite{asru11/kaldi}~$\blacklozenge$ + LM & -~~/ 8.6 & - & - \\
            Kaldi/chain~\cite{asru11/kaldi}~$\blacklozenge$ + LM & -~~/ 7.5 & - & - \\
            \midrule
            \multicolumn{4}{c}{\textit{Ours}} \\
            \midrule
            AR-Transformr~$\blacktriangle$ & 5.6 / 6.1 & 0.1703 & 29.7 M\\
            NAR-Transformer~$\blacktriangle$ & 5.6 / 6.3 & 0.0037 & 29.7 M\\
            AR-Transformer~$\blacktriangle\blacklozenge$ &  \textbf{5.2} / \textbf{5.7} & 0.1703 & 29.7 M\\
            NAR-Transformer~$\blacktriangle\blacklozenge$ & 5.3 / 5.9  & \textbf{0.0037} & 29.7 M\\
            \bottomrule[1.5pt]
        \end{tabular}
    }
    \vspace{-10pt}
    \label{tab:aishell-1-compare}
\end{table}

\subsection{1000-hours Corpus: AISHELL-2}
We then conduct our experiments on AISHELL-2~\cite{arxiv/aishell-2}, which is by far the largest free corpus available for Mandarin ASR research and contains 1000 hours of reading speech for training. As far as we know, none of the previous NAR methods have been tested on such a large dataset thus we are the first to evaluate the effectiveness of NAR models on industrial level Large Vocabulary Continuous Speech Recognition (LVCSR).

In Table~\ref{tab:aishell-2-compare}, it is worth noting that the state-of-the-art CER achieved by CIF~\cite{ic20/cif} model uses nearly three times more parameters than ours and the AR baseline we used is still competitive since it performs much better than other models. By comparing the results of AR and NAR transformer, we can draw a similar conclusion as in Section \ref{sec:aishell1} that our NAR model can achieve much faster decoding speed while maintaining the performance of its AR counterpart.

\begin{table}[!htp]
    \centering
    \vspace{-5pt}
    \caption{CER comparison with Hybrid and End-to-End models on 1000 hours of reading corpus Aishell-2~\cite{arxiv/aishell-2}. $\blacktriangle$: SpecAugment~\cite{is19/specmask} is used.
            $\blacklozenge$: Speed-Perturb~\cite{is15/speed_perturb} is used.
            $\blacktriangledown$: Adversarial-Data-Augmentation~\cite{taslp/Adversarial-egularization} is used. }
    \vspace{5pt}
    \scalebox{0.75}{
        \begin{tabular}{c|c|c|c}
            \toprule[1.5pt]
            Model  & IOS / Mic / Android & RTF & Params \\
            \midrule
            \multicolumn{4}{c}{\textit{End-to-End Autoregressive}} \\
            \midrule
            LAS~\cite{taslp/Adversarial-egularization}~$\blacktriangledown\blacklozenge$ & 9.2 / 10.3 / 9.7 & - & -\\
            S2S~\cite{is19/Ectc-Docd}~ + LM & 8.5 /~-~~-~~/~-~~- & - & - \\
            CIF~\cite{ic20/cif}~$\blacktriangle\blacklozenge$ + LM & 5.8 / 6.3 / 6.2 & - & 67 M + 22 M\\
            \midrule
            \multicolumn{4}{c}{\textit{Traditional Hybrid}} \\
            \midrule
            tri3(LDA+MLLT)~\cite{arxiv/aishell-2}~$\blacklozenge$ & 19.8 / 21.1 / 21.0 & - & -\\
            Chain-TDNN~\cite{arxiv/aishell-2}~$\blacklozenge$ & 8.8 / 10.9 / 9.6 & - & -\\
            \midrule
            \multicolumn{4}{c}{\textit{Ours}} \\
            \midrule
            AR-Transformer~$\blacktriangle\blacklozenge$ & 6.8 / 7.7 / 7.8 & 0.1703 & 29.7 M\\
            NAR-Transformer~$\blacktriangle\blacklozenge$ & 7.1 / 8.0 / 8.1 & \textbf{0.0037} & 29.7 M\\
            \bottomrule[1.5pt]
        \end{tabular}
    }
    \vspace{-10pt}
    \label{tab:aishell-2-compare}
\end{table}

\section{Conclusion}
\label{sec:conclusion}
In this paper, we proposed a novel Non-autoregressive Transformer with CTC-enhanced decoder input, which generates a sequence by refining the CTC prediction. The experimental comparisons demonstrate that on Aishell-1 and Aishell-2 tasks, our method can surpass all previous NAR models and approach the performance of AR models while achieving \textbf{50x} faster decoding speed.

\section{Acknowledgements}
\label{sec:ack}
This work was conducted when the first author was an intern at Tencent AI Lab, and is supported by National Natural Science Foundation of China (NSFC) (62076144) and joint research fund of NSFC-RGC (Research Grant Council of Hong Kong) (61531166002, N\_CUHK404/15). We would also like to thank Tencent AI Lab Rhino-Bird Focused Research Program (No. JR202041, JR201942) and Tsinghua University - Tencent Joint Laboratory for the support.

\vfill\pagebreak

\bibliographystyle{IEEEbib}
\bibliography{strings,refs}

\end{document}